\documentclass{article}
\usepackage{spconf,amsmath,graphicx}

\usepackage{cite}
\usepackage{amsmath,amssymb,amsfonts}
\usepackage{algorithmic}
\usepackage{graphicx}
\usepackage{textcomp}
\usepackage{xcolor}
\def\BibTeX{{\rm B\kern-.05em{\sc i\kern-.025em b}\kern-.08em
    T\kern-.1667em\lower.7ex\hbox{E}\kern-.125emX}}
\usepackage{subcaption}

\title{J-SGFT: Joint Spatial and Graph Fourier Domain Learning for Point Cloud Attribute Deblocking}

\usepackage[symbol]{footmisc}

\name{%
  Muhammad Talha\textsuperscript{1\thanks{This work is supported by the NSF award [CNS-2148382].}},%
  Qi Yang\textsuperscript{1},%
  Zhu Li\textsuperscript{1},%
  Anique Akhtar\textsuperscript{2},%
  Geert Van Der Auwera\textsuperscript{2}%
}
\vspace{-4mm}
\address{%
  \textsuperscript{1}University of Missouri–Kansas City\\
  \textsuperscript{2}Qualcomm Technologies
}
\vspace{-5mm}

\begin{document}
%
\maketitle
\vspace{-6mm}
\begin{abstract}

Point clouds (PC) are essential for AR/VR and autonomous driving but challenge compression schemes with their size, irregular sampling, and sparsity. MPEG’s Geometry-based Point Cloud Compression (GPCC) methods successfully reduce bitrate; however, they introduce significant blocky artifacts in the reconstructed point cloud. We introduce a novel multiscale postprocessing framework that fuses graph-Fourier latent attribute representations with sparse convolutions and channel-wise attention to efficiently deblock reconstructed point clouds. Against the GPCC TMC13v14 baseline, our approach achieves BD rate reduction of 18.81\% in the Y channel and 18.14\% in the joint YUV on the 8iVFBv2 dataset, delivering markedly improved visual fidelity with minimal overhead.

\end{abstract}
\vspace{-5mm}
\section{Introduction}
\label{sec:intro}
\vspace{-4mm}

Point clouds provide precise 3D data for industrial robotics and autonomous navigation \cite{8825788}, but their millions of nonuniform points challenge storage, management, and traditional compression/filtering \cite{li2019puganpointcloudupsampling}. To reduce size, Geometry-based Point Cloud Compression (GPCC) \cite{9457097} and Video-based Point Cloud Compression (VPCC) \cite{Graziosi_Nakagami_Kuma_Zaghetto_Suzuki_Tabatabai_2020} are employed; this paper focuses on GPCC. Although GPCC achieves high compression ratios, it can introduce noise and quantization artifacts in both geometry and attributes \cite{li2019puganpointcloudupsampling}, degrading visual quality and reconstruction accuracy in high-fidelity applications. Therefore, eliminating lossy compression distortions is critical and valuable. Despite extensive geometry-based deblocking research, attribute-based deblocking remains underexplored, which is the main topic of this paper. To the best of our knowledge, there are five methods that directly address this challenge: MS-GAT \cite{9767661}, CARNet \cite{ding2022carnetcompressionartifactreductionpoint}, MUSCON \cite{10533833}, TSFNet3d \cite{10648101}, and \cite{10743459}. MS-GAT employs graph Laplacians, Chebyshev convolutions, and attention mechanism to capture irregularities in spatial domain but yields only marginal gains at high computational cost. CARNet demonstrates better results by using sparse convolutions and MLPs to separate high- and low-frequency signals, yet its average pooling of high-frequency components still induces information loss. MUSCON pioneers multiscale attribute upsampling with SparseConv, fusing fine and coarse representations from multiple scales and delivers far superior deblocking performance. However, MUSCON’s reliance on feature averaging during downsampling induces information loss, which ultimately gives room for improvement.
TSFNet3d \cite{10648101}, while achieving state-of-the-art results, similarly averages features across three scales, suffering the same drawback. These limitations highlight the need for an even more
efficient method for attribute-based deblocking. However, recent works like \cite{shao2017attributecompression3dpoint,pavez2020regionadaptivegraphfourier, 10743459} have demonstrated that since point clouds are mostly irregular and sparse in nature, interpreting them as graph signals has proven to be highly effective for both reconstruction and compression efficiency.
Thus, we introduce \textbf{Joint Spatial and Graph Fourier Domain Learning for Point Cloud Attribute Deblocking (J-SGFT)} that extends multi-scale feature learning approach to frequency domain utilizing graph fourier transform (GFT) tools. By eschewing feature averaging—and thus eliminating any attendant information loss—our approach delivers markedly superior performance compared to all prior methods. 
\begin{figure*}[ht]
    \centering
    \includegraphics[width=\textwidth]{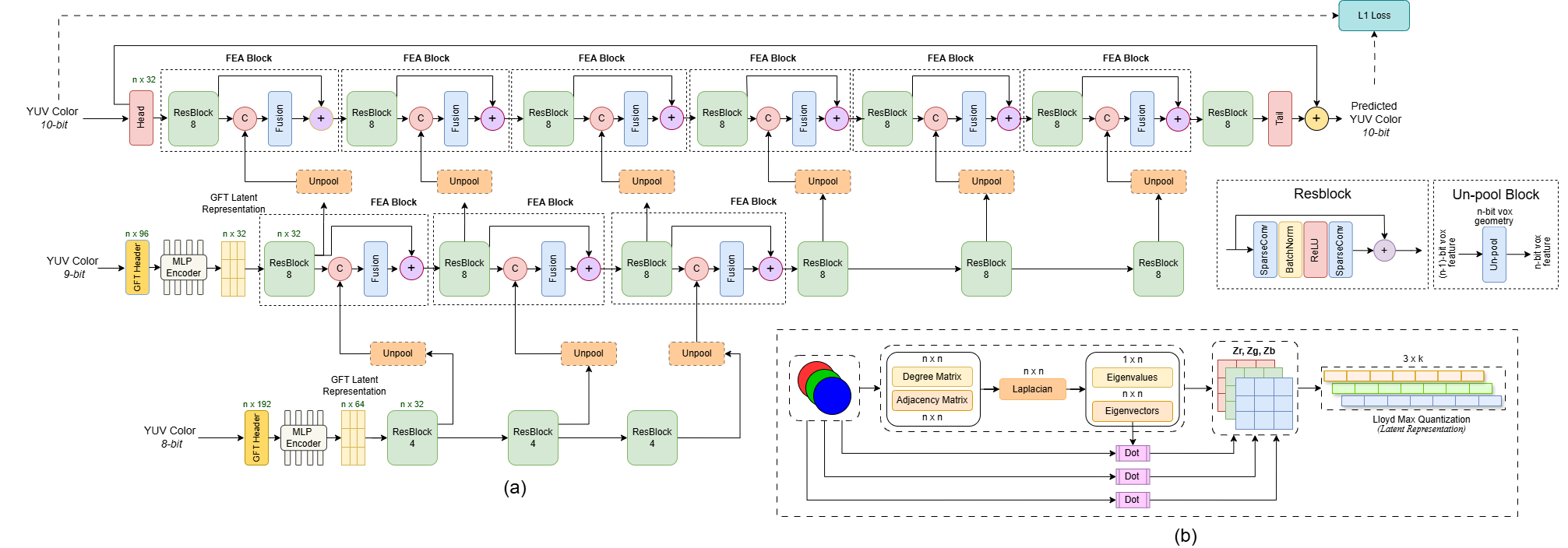}
    \vspace{-6mm}
    \caption{(a) Shows J-SGFT: Multi-scale Architecture for deblocking. (b) GFT Header: Input to this block is color attributes \textbf{\textit{(Y, U, V)}}, where \textbf{Zr, Zb, Zb} are attributes in Graph fourier domain which are further quanztized into \textbf{\textit{n}} number of bins to get latent representation.}
    \label{fig:main_architecture}
    \vspace{-5mm}
\end{figure*}

Our main contributions are as follows:
\vspace{-4mm}

\begin{enumerate}
    \setlength{\itemsep}{0pt} 
    \setlength{\parsep}{0pt}  
    \setlength{\topsep}{0pt}  
    \item We propose J-SGFT, the first joint spatial and spectral domain learning solution for point cloud attribute (PCA) deblocking. We develop a GFT Header to project PCA signals to the spectral domain.

    \item Following GFT Header, Lloyd Max's algorithm to find latent representation at low resolution (LR) which is sent to an MLP that reduces its dimensions to control computational cost. 

    \item Multi-scale SparseConv-based learning is used to leverage PC sparsity to extract and refine latent features at multiple resolution. A channel-wise attention block is then proposed to fuse LR features into the high resolution branch, optimizing the reconstruction quality.

\end{enumerate}
\vspace{-2mm}
The experiment results show that J-SGFT achieves a substantial BD-rate reduction of 18.13\%, outperforming the state-of-the-art CARNet, which offers a 9\% BD-rate reduction for YUV-optimized models.

\vspace{-6mm}
\section{JOINT SPATIAL AND SPECTRAL ATTRIBUTE LEARNING}
\label{sec:method}
\vspace{-3mm}

Inspired by MUSCON, our work consists of a sophisticated multi-scale architecture, which can handle data at three different resolutions as Fig. \ref{fig:main_architecture}. The architecture consists of: (1) a GFT header to convert features into the frequency domain, (2) the Lloyd-Max algorithm for adaptive binning and latent representation, (3) a small MLP for latent dimensionality reduction, (4) a channel-wise attention mechanism for feature fusion, and (5) a multi-scale architecture that jointly learns spatial and GFT-based representations. To avoid information loss casued by downsampling, we integrate GFT within a multi-scale framework, preserving high-resolution (HR) details by projecting attributes into the frequency domain, capturing both AC and DC components. Unlike traditional methods that average attributes, our approach maps frequency responses to lower-resolution voxels, retaining critical information. Specially, Two lower resolutions are generated from 10-bit, i.e., 9-bit and 8-bit. 10-bit branch works in spatial domain (Y,U,V) \cite{7025414} and other two resolutions are convert to latent representation with GFT header and ResBlock. 
\vspace{-5mm}
\subsection{Graph Fourier Transform Header}
\label{ssec:GFThead}
\vspace{-2mm}

\begin{figure*}[htbp]
    \centering
    \begin{minipage}{0.24\textwidth}
        \centering
        \includegraphics[width=\textwidth]{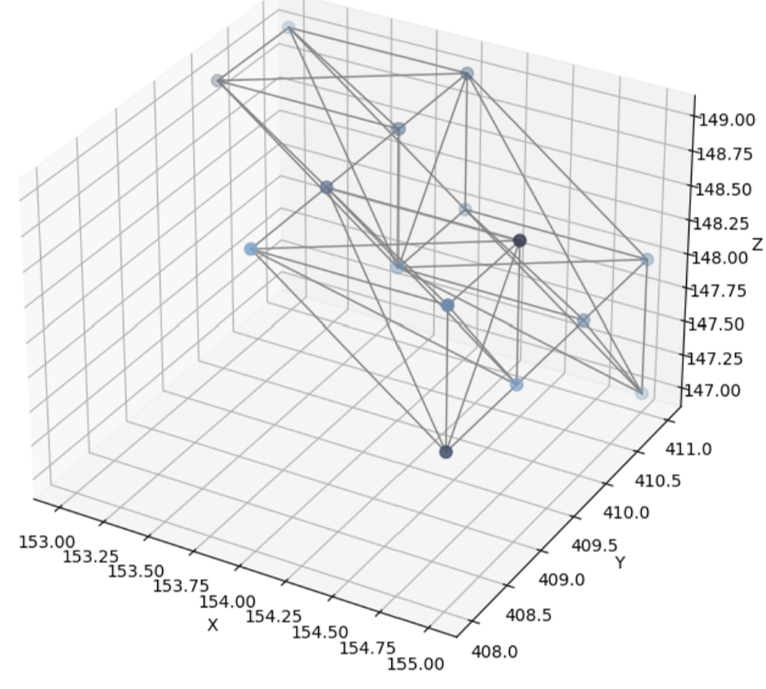}
        \subcaption{(a)}\label{fig:graph_1}
    \end{minipage}%
    \begin{minipage}{0.24\textwidth}
        \centering
        \includegraphics[width=\textwidth]{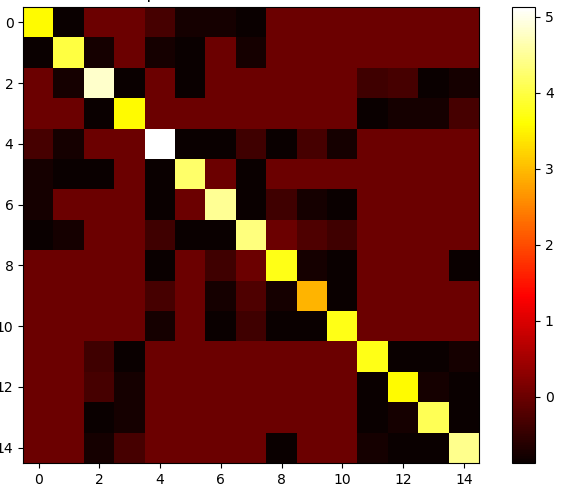}
        \subcaption{(b)}\label{fig:2}
    \end{minipage}%
    \begin{minipage}{0.48\textwidth}
        \centering
        \includegraphics[width=\textwidth]{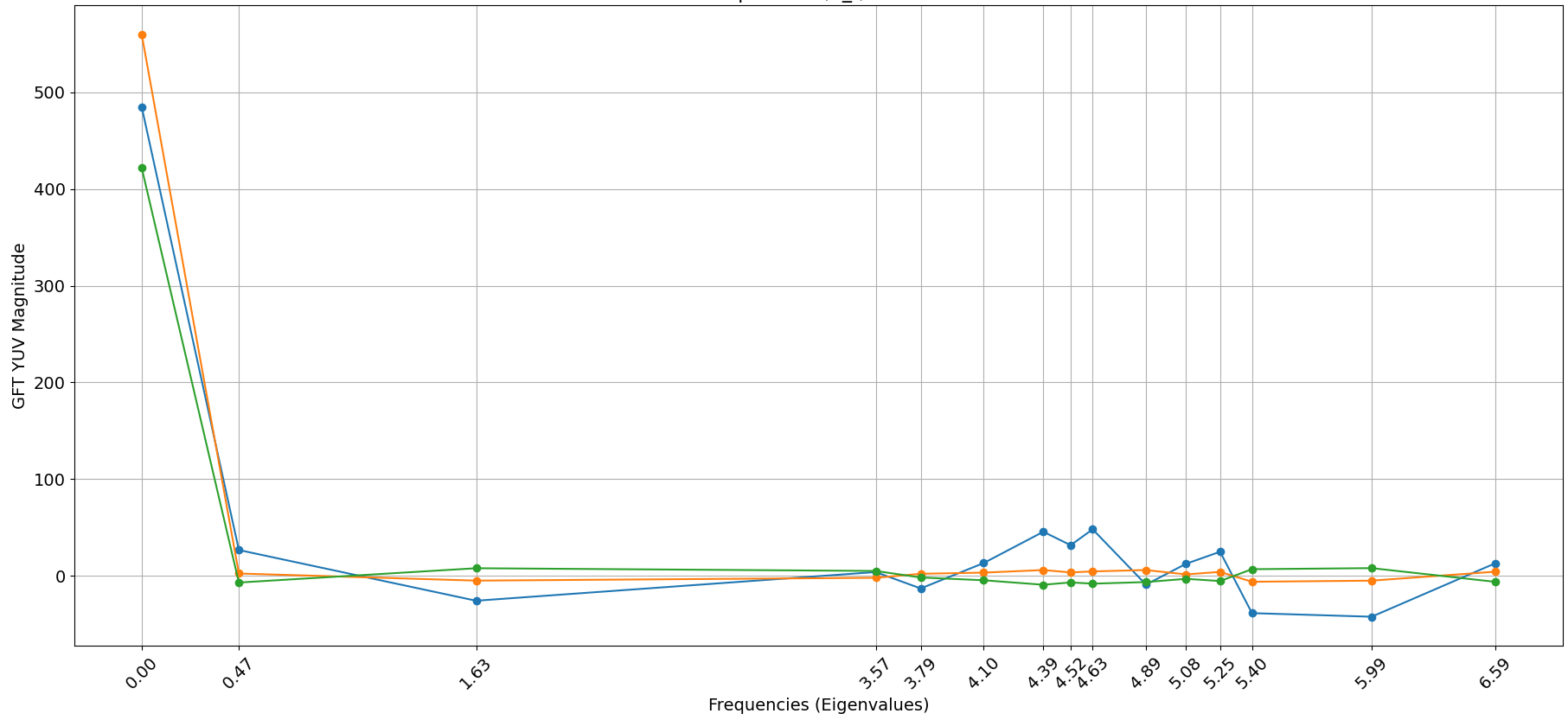}
        \subcaption{(c)}\label{fig:3}
    \end{minipage}%
    \label{fig:graph}
    \vspace{-3mm}
\end{figure*}

\begin{figure*}[htbp]
    \centering
        \centering
        \includegraphics[width=\textwidth]{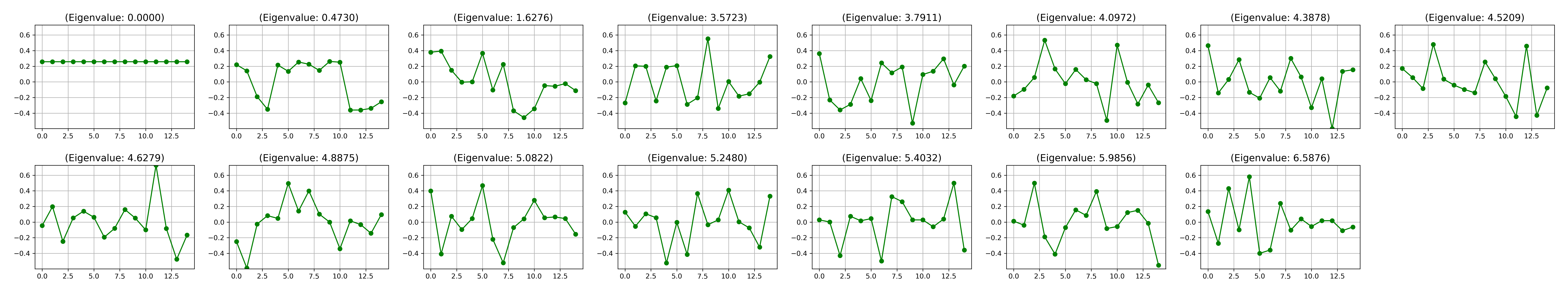}
        \label{fig:graph_2}
        \vspace{-3mm}
        \subcaption{(d)}
        \caption{Graph Fourier Transform: (a) shows graph in 4 × 4 × 4 voxel in Longdress sequence. (b) Displays the graph connectivity between nodes via Laplacian matrix. (c) gives the eigenvalues associated with the signal interpreted as signal frequencies, and shows YUV signal projected in spectral domain as Z(f) against those eigenvalues. (d) Eigenvectors demonstrating the variation trends against their corresponding eigenvalues.}
        
    \label{fig:graph}
    \vspace{-5mm}
\end{figure*}

Point cloud data are inherently sparse and irregular, making them well-suited for representation as graphs. In this context, each point is treated as a node, with edges encoding spatial, temporal, or feature-based relationships. By converting point clouds into graphs, we can leverage graph-based techniques to extract the underlying 3D structure and inter-dependencies from frequency domain, addressing their inherent sparsity and irregularity. To map the point cloud attributes to the frequency domain, we first organize the points using a k-d tree, which partitions the data into voxel blocks through recursive binary splitting, as described in \cite{10.1145/361002.361007}. Within each voxel, we construct a graph by connecting points as nodes, with edges representing spatial or feature-based relationships as Fig. \ref{fig:graph}a shown. This graph structure is formulated as: $\textit{G} = (\nu ,\Xi)$. where $\nu$ represents the set of nodes (vertices) in the graph and is defined as: $\nu = \{n_1, n_2, ....., n_i \} $, and $ \Xi$ shows the set of edges $\Xi$ = $ \{ \xi_1, \xi_2,...., \xi_j \} $. $\xi_j$ is a binary indicator that denotes an edge connecting two points $n_p$ and $n_q$. The first step in applying the GFT \cite{pavez2020regionadaptivegraphfourier} is computing the non-normalized graph Laplacian: $\mathcal{L} = D - A$ in which $D$ and $A$ represent degree and adjacency matrix.

The adjacency matrix $A$ has entries $\omega_{i,j}$, 
$w_{ij} = e^{-\alpha \cdot d_{ij}}$
where $d_{ij}$ denotes the Euclidean distance between the color values of points $n_i$ and $n_j$ and $\alpha$ is a scaling factor that controls the rate at which the weight decreases with increasing distance,  representing the weight of the edge between nodes $n_i$ and $n_j$. These weights capture the strength of connectivity and influence the graph spectrum, determining how information propagates through the network. The degree matrix $D$ is a diagonal matrix that encodes the connectivity degree of each node within a block, given by 

 \vspace{-2mm}
\begin{equation} 
D_i = \sum_{j} w_{ij}
 \vspace{-3mm}
 \label{eq:degree}
\end{equation}

Solving the eigenproblem $ \mathcal{L}\,\mu = \lambda\,\mu$
yields eigenvalues \(\lambda\) (the signal’s frequencies) and the orthonormal eigenvector matrix \(\mu\), whose mutual orthogonality and normalization ensure a stable, reversible transform. As in standard Fourier analysis, small \(\lambda\) correspond to low-frequency (smooth, global) variations, while large \(\lambda\) capture high-frequency (fine, localized) details. Fig.~\ref{fig:graph}d illustrates the evolution of eigenvectors with their eigenvalues. We apply \(\mu\) in each block to map spatial attributes \(\iota\) (the color components Y, U, V) to the frequency domain: $\mathcal{Z}\,\iota = \mu \cdot \iota $, where \(\mathcal{Z}\,\iota\) is the resulting frequency response. All intermediate matrices (adjacency, Laplacian, eigenvector) are \(n\times n\) for a voxel with \(n\) points; both \(\lambda\) and \(\mathcal{Z}\,\iota\) are \(1\times n\) per attribute. Figs.~\ref{fig:graph}a and \ref{fig:graph}c show the spatial and GFT latent representations, respectively, of a \(4\times4\times4\) voxel in the longdress sequence, with Fig.~\ref{fig:graph}c interpreting the eigenvalues as frequencies.

\vspace{-5mm}
\subsection{Latent Representation}
\label{ssec:latent}
\vspace{-2mm}

We map the color attributes projected in frequency domain, i.e., (\(Z_y\), \(Z_u\), \(Z_v\)) to \(K\) discrete bins using Lloyd-Max quantization \cite{8682396}. Centroids, uniformly initialized over \([-f_{\text{max}}, f_{\text{max}}]\), are iteratively refined to minimize the mean-squared error between the eigenvalues and their centroids; each eigenvalue is then assigned to its nearest centroid. The three attributes are quantized independently:
\textbf{9-bit branch} \(2 \times 2 \times 2\) voxel: each attribute  yields \(n \times 32\)  coefficients, giving a latent tensor of  \(n \times 3 \times 32\) = \(n \times 96\).
\textbf{8-bit branch} \(4 \times 4 \times 4\) voxel: each attribute yields  \(n \times 64\) coefficients, giving a latent tensor of \(n \times 3 \times 64\) = \(n \times 192\). Within every frequency bin, any array-valued color entries are summed to scalars before assignment. The resulting per-voxel, quantized frequency-domain representations are stored and passed to the deblocking model. The color attributes in the frequency domain can thus be expressed as:
\vspace{-4mm}
\begin{equation} 
\label{eq:eigen}
\mathcal{Z}_y = \sum_{i=1}^k \mathbb{1}_{\{\text{bin\_index}(f_i) = i\}} \cdot Z_y^{(i)}
\vspace{-3mm}
\end{equation}

\vspace{-3mm}

\begin{figure}[ht]
    \centering
    \scalebox{0.5}{\includegraphics{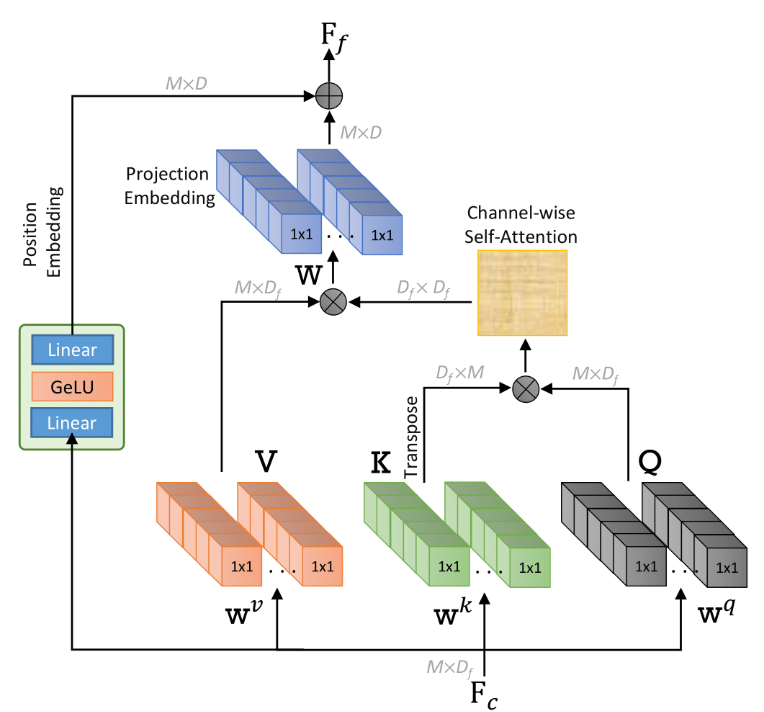}}
    \vspace{-5mm}
    \caption{Channel-wise attention based fusion block}
    \label{fig:channel-wise-attention}
    \vspace{-6mm}
\end{figure}

Here, \(Z_y^{(i)}\) is the quantized Y-channel value for points in bin \(i\), with \(\mathbb{1}\) as the indicator function. The U- and V-channel quantized representations, \(\mathcal{Z}_u\) and \(\mathcal{Z}_v\), are defined analogously.

\vspace{-6mm}

 \subsection{MLP for Dimension Reduction}
\label{ssec:MLP}
\vspace{-2mm}

To manage the high-dimensional latent outputs of the GFT Header, we employ a small UNet–style MLP encoder–decoder. We train it on both 9-bit and 8-bit latent representations, then fix the encoder to perform dimensionality reduction: in the 9-bit branch, it compresses \(n \times 96\) to \(n \times 32\); in the 8-bit branch, it reduces \(n \times 192\) to \(n \times 64\).
\vspace{-5mm}
\subsection{Feature Extraction and Aggregation}
\label{sssec:FEA}
 \vspace{-2mm}
As seen in Fig. \ref{fig:main_architecture}a, every branch in our framework has cascades of Feature Extraction and Aggregation (FEA) blocks. Each FEA block takes two inputs. The main input to this FEA block is a sparse tensor $\zeta$, from the current n-bit branch, and a secondary input is the upscaled features from the (n-1)th-bit branch which are then concatenated together for better feature understanding. FEA block is divided into three components: Resblock, concatenation, and Fusion. ResBlock is structured with a sequence of layers, represented as \textit{\textbf{“conv $\rightarrow$ BN $\rightarrow$ ReLU $\rightarrow$ conv”}}. We apply a direct skip connection from the input of the first convolutional layer to the output of the last convolutional layer within the ResBlock to deal with the vanishing gradient problem as Fig. \ref{fig:main_architecture}. Network activations are normalized by the batch normalization layer for a stable learning process. 
\vspace{-4mm}
\subsection{Feature Aggregation}
\label{sssec:fusion}
\vspace{-2mm}
In the second half of the FEA block, the focus shifts to feature aggregation, which involves merging and integrating feature maps. Inspired by \cite{10275101}, we introduce a channel-wise attention mechanism for feature fusion, as shown in Fig.~\ref{fig:channel-wise-attention}. This mechanism plays a crucial role in fusing feature sets from multiple branches, enhancing the representational power of multidimensional data. It employs linear projections to transform the input feature matrix, $F_c \in \mathbb{R}^{M \times D_f}$, into three components—queries $(Q)$, keys $(K)$, and values $(V)$—each defined by their respective weight matrices $W^Q, W^K, W^V \in \mathbb{R}^{D_f \times D_f}$. The transformation process is formalized as follows:

\vspace{-5mm}
\begin{equation} 
Q = F_c W^Q,  K = F_c W^K,  V = F_c W^V.
\label{eq:channel-wise}
\vspace{-1mm}
\end{equation}
These transformations yield matrices that encapsulate the essence of the feature space in a format conducive to attention-based operations. Subsequently, the attention scores are computed through the transposition and multiplication of the matrices $K$ and $Q$, and a softmax normalization is applied to this product, ensuring an equitable distribution of attention across the channels, Eq. \ref{eq:channel-wise}.
This block is the key to combining separately processed features from different branches. 
\vspace{-1mm}
\begin{equation} 
\vspace{-2mm}
A = softmax(K^T \cdot Q)
\label{eq:channel-wise-atten}
\vspace{-1mm}
\end{equation}

Matrix \(A\) encodes feature importance, weighting the value matrix \(V\) to amplify relevant features and suppress less pertinent ones, thereby ensuring contextually enriched representations that respect the spatial and feature-specific dependencies of the data. The network then fuses these attention-weighted features with the positionally embedded features to produce an output feature matrix \(F_f \in \mathbb{R}^{M \times D}\), where \(D\) is the feature dimension post-fusion. This channel-wise self-attention–guided fusion synthesizes a comprehensive, context-aware representation of the input, marking a significant advancement in feature fusion methodologies.

\vspace{-4mm}
\subsection{Unpool Block}
\label{ssec:unpool}
\vspace{-2mm}
The Unpool block aligns the \((n-1)\)-bit and \(n\)-bit geometries before merging features in the Fusion block, enabling concatenation across dual branches, Fig. \ref{fig:main_architecture}. As shown in \cite{7410535}, it combines the \((n-1)\)-bit and \(n\)-bit features to generate voxelized representations, which are essential for SparseTensor conversion and compatibility within the architecture. The voxelization, as defined in Eq.~\ref{eq:downsample}, downscales the \(n\)-bit point cloud to the \((n-1)\)-bit format, preserving key geometric properties:

\vspace{-3mm}
\begin{equation}
 \vspace{-3mm}
C_{n-1} = \text{round}\left( \left( 2^{n-1} - 1 \right) \cdot \frac{C_n}{2^n - 1} \right)
\label{eq:downsample}
\end{equation}
\vspace{-1mm}

\begin{figure*}[htbp]
    \centering
    \includegraphics[width=\linewidth]{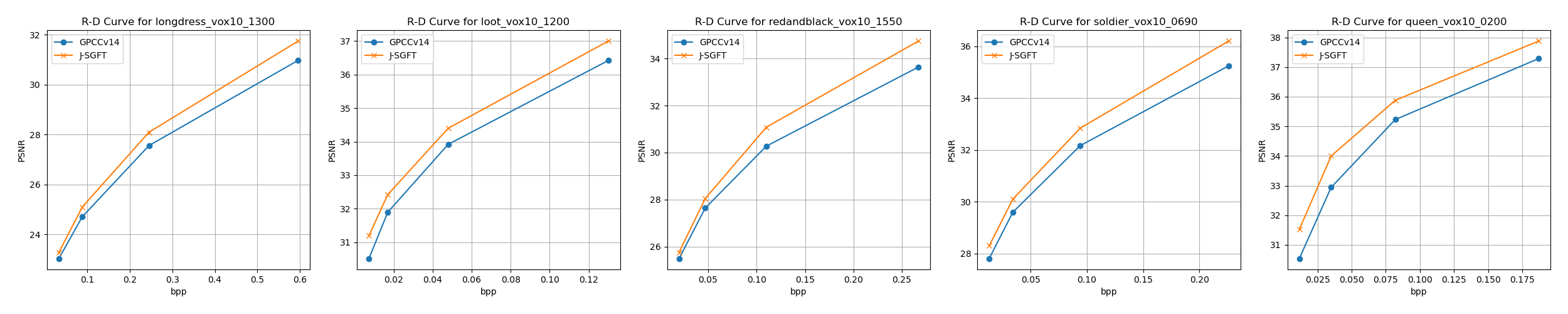}
        \vspace{-8mm}
    \caption{RD-curve for Y-PSNR comparison between our method and GPCC TMC13v14.}
    \label{fig:RD Curve}
    \vspace{-4mm}
\end{figure*}
In our architecture, voxelization begins with a 10-bit HR point cloud, which is progressively downscaled to 9-bit and 8-bit representations, maintaining geometric consistency. These multi-resolution datasets are then processed by three specialized branches corresponding to 10-bit, 9-bit, and 8-bit depths. The 10-bit branch starts with a head module that uses SparseConv layers and increases the feature size to 32, while the 9-bit and 8-bit branches incorporate a GFT head for frequency domain pre-processing (See. \ref{ssec:GFThead}). The 8-bit branch employs two Unpool blocks to upscale the point cloud to 9-bit and 10-bit representations, followed by ResBlocks for further feature refinement. FEA blocks in the 9-bit and 10-bit branches enhance multiscale information. The outputs from all three branches are passed through ResBlocks in the tail layer, which reduces the features to a 3-channel format corresponding to the YUV color space. We evaluate the network's performance using BD-rate and PSNR metrics.

We use L1 loss that guarantees a thorough optimization. However, the $L1$ loss is computed for Y, U, and V components individually as $\mathcal{L}_{Y}$, $\mathcal{L}_{U}$, and $\mathcal{L}_{V}$, respectively. Then, the final loss $\mathcal{L}_{joint}$ is obtained using Eq. \ref{eq:eq_2}.
\vspace{-2mm}
\begin{equation} \label{eq:eq_2}
    \mathcal{L}_{joint} = \frac{6}{8}\mathcal{L}_{Y} + \frac{1}{8}\mathcal{L}_{U} + \frac{1}{8}\mathcal{L}_{V}
\end{equation}

\vspace{-7mm}
\section{TRAINING $\&$ IMPLEMENTATION}
\label{training}
 \vspace{-3mm}
\subsection{Training / Validation $\&$ Testing Dataset}

We train on the THUman2.0 corpus, comprising \(526\) full-body meshes \cite{9577865}. From each mesh we uniformly sample \(5\times10^{6}\) surface points, voxelizing them into \(10\)-bit grids that each contain \(\sim10^{6}\) points; colours are interpolated by the nearest original surface point. The first \(511\) clouds form the training set and the remaining \(15\) the validation set. All point clouds are compressed with the GPCC codec TMC13v14 \cite{mpeg} under the RAHT profile (lossless geometry, lossy attributes) \cite{7482691}. We repeated every experiment with TMC13v21 and obtained comparable outcomes, but we report TMC13v14 for parity with prior work (CARNet, MS-GAT); because CARNet markedly outperforms MS-GAT, we restrict quantitative comparisons to CARNet. Each cloud is partitioned into \(16\) patches using a depth-\(4\) binary tree. Patch geometry is downsampled via Eq.~\ref{eq:downsample} to produce \(9\)-bit and \(8\)-bit versions. Rather than averaging attributes, we convert each voxel’s colour to its spectral representation and remap these coefficients to the spatial positions of the downsampled cloud, preserving all information. This yields \(8\,191\) training patches and \(236\) validation patches. For testing we employ the 8iVBFv2 dataset, compressed with identical codec settings so that its spatial structure and latent attributes mirror those in training.

\begin{table}[h!]
\centering
\vspace{-5mm}
\caption{Y-BD rate comparison between our model and MUSCON \cite{MUSCON}. YUV-BD rate (Joint-Model) comparison between our model, CARNet, and TSFNet3d}
\vspace{-3mm}
\footnotesize 
\resizebox{\columnwidth}{!}{ 
\setlength\tabcolsep{2pt} 
\begin{tabular}{l|c|c|c|c|c}
\hline
\multicolumn{6}{c}{\textbf{TMC13v14}} \\ \hline
& \multicolumn{2}{c|}{\textbf{Y-BD Rate}} 
& \multicolumn{3}{c}{\textbf{YUV-BD Rate}} \\ \hline
\textbf{Sequences} & \textbf{Ours} & \textbf{MUSCON} & \textbf{Ours} & \textbf{CARNet} & \textbf{TSFNet3d} \\ \hline
Longdress & \textbf{-15.0651} & -12.96 & \textbf{-13.66} & -9.05 & -13.61 \\ \hline
Loot & \textbf{-18.67} & -16.54 & \textbf{-18.5} & -5.72 & -17.61 \\ \hline
RedandBlack & \textbf{-18.5614} & -16.48 & -17.3 & -8.35 & \textbf{-19.8} \\ \hline
Soldier & \textbf{-18.7327} & -17.26 & \textbf{-19.61} & -4.51 & -19.3 \\ \hline
Queen & \textbf{-23.0699} & -17.26 & \textbf{-21.62} & -13.78 & -15.72 \\ \hline
\textbf{Average} & \textbf{-18.81} & \textbf{-15.81} & \textbf{-18.138} & -8.282 & -17.208 \\ \hline
\end{tabular}
}
\label{table:bdrate_com}
\vspace{-3mm}
\end{table}

\vspace{-2mm}
\begin{figure}[ht]
    \centering
    \includegraphics[width=\linewidth]{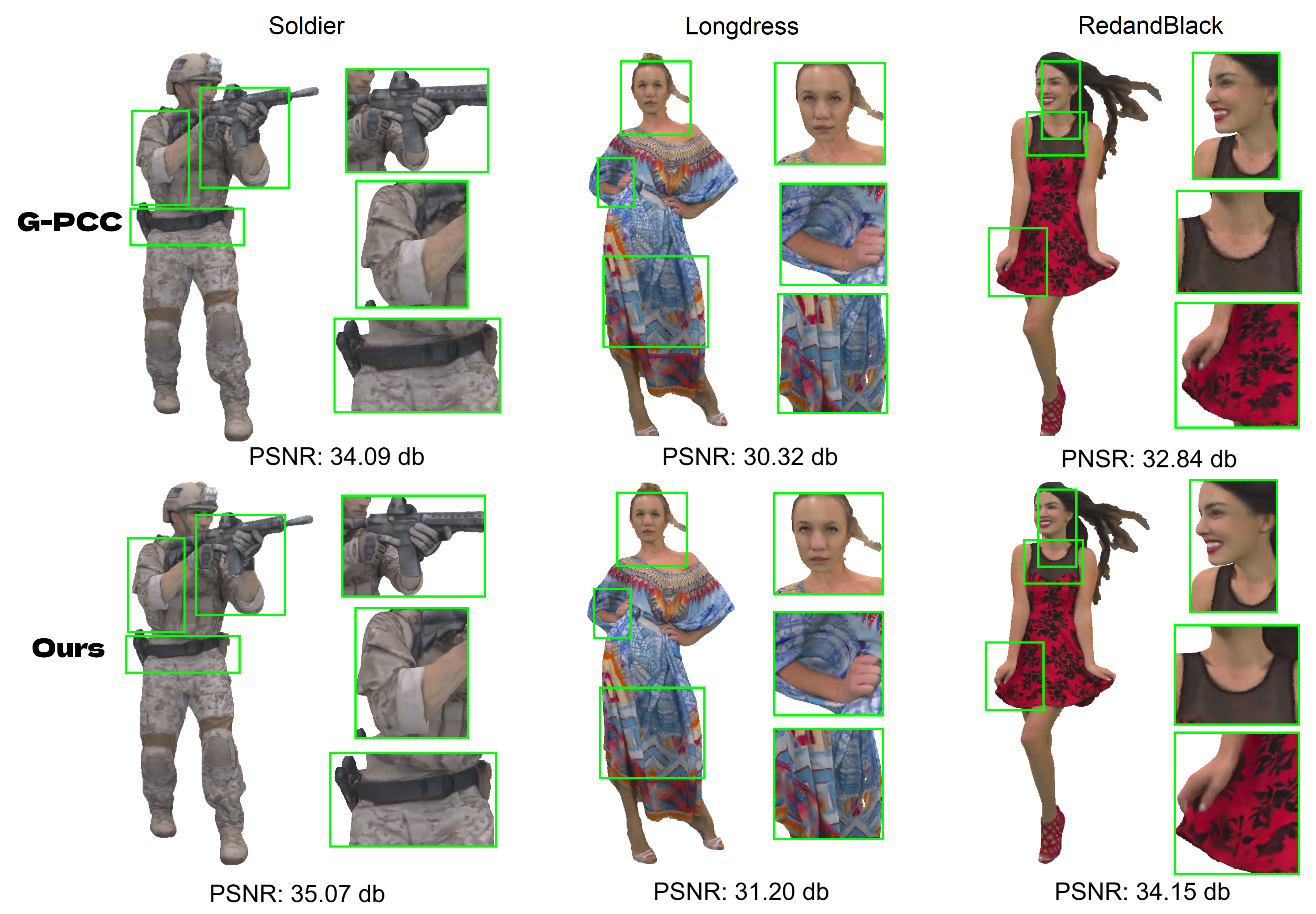}
     \vspace{-4mm}
    \caption{Visual Comparison of GPCCv14-encoded and our method's predicted point cloud on rate 4.}
    \label{fig:result}
    \vspace{-5mm}
\end{figure}

\vspace{-3mm}

\subsection{Training Implementation}
\vspace{-3mm}
We base our work on the Minkowski Engine in PyTorch \cite{8953494} due to its efficiency with sparse tensors and irregular 3D shapes. The network is optimized using the Adam optimizer with $\beta_1 = 0.9$ and $\beta_2 = 0.999$. For deblocking, the learning rate starts at $1 \times 10^{-4}$ and is reduced to $1 \times 10^{-6}$ using a cosine-annealing scheduler. The model is trained for 60 epochs with a batch size of 2, totaling 5.6 million parameters. 

The GFT Header has a theoretical complexity of $O(n^3)$, but since operations are performed at the voxel level with only a few points, the graph size remains small, resulting in per-voxel computation times less than milliseconds.
\vspace{-5mm}
\subsection{Experiment Results}
\label{ssec:results}
\vspace{-2mm}
To evaluate the model's performance, we follow MPEG's standard test conditions, using \textit{bpp} (bits per point) and \textit{Y-PSNR} (Peak Signal-to-Noise Ratio) to assess data efficiency and the quality of the Y-component reconstruction. We compute average rate-distortion performance using the Bjøntegaard delta bit rate (BD-rate) across four quantization parameters. We compare our proposed network with the GPCC TMC13v14 and the sparseConv-based MUSCON \cite{MUSCON}. Our model achieves a 18.81\% BD-rate reduction over GPCCv14 (Table~\ref{table:bdrate_com}) and demonstrates superior Y-PSNR performance, as shown by the RD-Curve in Fig. \ref{fig:RD Curve}. Additionally, it also outperforms MUSCON on the 8iVFBv2 dataset (using the same test sequences: 'longdress', 'Loot', 'Soldier', 'Redandblack') which gives 15.81\% bd-rate reduction. 
Table~\ref{table:bdrate_com} also shows the comparison of YUV-BD rate between our proposed work, CARNet and TSFNet3D. As we trained YUV channels together, we compare its results with Joint model performance of CARNet and TSFNet3d. It is evident that proposed \textbf{J-SGFT} outperforms CARNet across all 8i sequences by a huge margin. It also achieves better results than TSFNet3D, except for only 'RedandBlack.' Notably, we achieve the best BD rate for 'Longdress' (-13.66\%), 'Loot' (-18.5\%), 'Soldier' (-19.61\%), and 'Queen' (-21.62\%), with an average improvement of -18.13\%, surpassing CARNet’s -8.28\% and TSFNet3D’s -17.21\%. These results demonstrate our model's robustness and superior performance in YUV space. Fig.~\ref{fig:result} shows the reconstructed point cloud attribute samples processed by both G-PCC and our method. G-PCC exhibits significant compression artifacts, such as blockiness and blurriness, especially on sequences like 'longdress' and 'RedandBlack.' In contrast, our method delivers superior visual quality, with blocky regions smoothed out and features closely resembling the ground truth. Trained on the THUman2.0 dataset and tested on 8iVFBv2, our approach generalizes well across datasets, demonstrating its robustness.
\vspace{-9mm}
\section{Conclusion}
\label{sec:Conclusion}
\vspace{-3mm}
We present a novel solution to the limitations of MPEG's TMC13v14, which, while effective for compressing geometry and attributes, introduces coding noise. Our approach leverages a multi-scale SparseConv-based architecture enhanced with Graph Fourier Transform (GFT) for latent representation, improving attribute learning and deblocking in point clouds. The key innovation in proposed work is our Joint Spatial and Spectral feature learning process, where GFT captures finer and coarser attribute signal information in spectral domain, integrated within a multi-resolution network using SparseConv to manage data sparsity. A channel-wise transformer fuses features from lower resolution branches to higher resolution branches that further enhances reconstruction quality. Our method achieves a \textbf{18.81\%} BD rate on Y-channel and \textbf{18.138\%} BD-rate reduction on YUV-channel beating the previous state-of-the-arts, setting a new benchmark in point cloud deblocking and reconstruction.

\bibliography{refs}

\end{document}